\documentclass[
aps,
prb,
showpacs,
floatfix,
reprint,
twoside,
superscriptaddress
]{revtex4-1}
\usepackage{amsmath}
\usepackage[T1]{fontenc}
\usepackage{fourier}
\usepackage{graphicx}
\usepackage[colorlinks=true,
allcolors=blue,
dvipdfm=true]{hyperref}

\begin{document}

\title{Josephson effect in multi-terminal superconductor-ferromagnet junctions coupled via triplet components}

\author{Andreas Moor}
\affiliation{Theoretische Physik III, Ruhr-Universit\"{a}t Bochum, D-44780 Bochum, Germany}
\author{Anatoly F.~Volkov}
\affiliation{Theoretische Physik III, Ruhr-Universit\"{a}t Bochum, D-44780 Bochum, Germany}
\author{Konstantin B.~Efetov}
\affiliation{Theoretische Physik III, Ruhr-Universit\"{a}t Bochum, D-44780 Bochum, Germany}
\affiliation{National University of Science and Technology ``MISiS'', Moscow, 119049, Russia}

\begin{abstract}
On the basis of the Usadel equation we study a multi-terminal Josephson junction. This junction is composed by ``magnetic'' superconductors~S$_{\text{m}}$ which have singlet pairing and are separated from the normal n~wire by spin filters so that the Josephson coupling is caused only by fully polarized triplet components. We show that there is no interaction between triplet Cooper pairs with antiparallel total spin orientations. The presence of an additional singlet superconductor~S attached to the n~wire leads to a finite Josephson current~$I_{\text{Q}}$ with an unusual current--phase relation. The density of states in the n~wire for different orientations of spins of Cooper pairs is calculated. We derive a general formula for the current~$I_{\text{Q}}$ in a multi-terminal Josephson contact and apply this formula for analysis of two four-terminal Josephson junctions of different structures. It is shown in particular that both the ``nematic'' and the ``magnetic'' cases can be realized in these junctions. In a two-terminal structure with parallel filter orientations and in a three-terminal structure with antiparallel filter orientations of the ``magnetic'' superconductors with attached additional singlet superconductor, we find a nonmonotonic temperature dependence of the critical current. Also, in these structures, the critical current shows a Riedel peak like dependence on the exchange field in the ``magnetic'' superconductors. Although there is no current through the S/n~interface due to orthogonality of the singlet and triplet components, the phase of the order parameter in the superconuctor~S is shown to affect the Josephson current in a multi-terminal structure.
\end{abstract}

\date{\today}

\maketitle

\section{Introduction}

During the last decade, there has been an increasing interest in studying the Josephson effect in Josephson junctions~(JJ) of different types. Interesting effects have been observed in JJs consisting of superconductors~(S) and ferromagnetic layers~(F). The authors of Refs.~\onlinecite{Bulaev77,Buzdin82} have predicted long ago that in JJs of the S/F/S~type, the critical current~$I_{\text{c}}$ may change sign and the so-called $\pi$\nobreakdash-state can be realized. However, only recently the sign reversal of~$I_{\text{c}}$ has been observed experimentally.\cite{Ryazanov_et_al_2001,Kontos_et_al_2002,Blum_et_al_2002,Bauer_et_al_2004,Sellier_et_al_2004,Shelukhin_et_al_2006,Bannykh_et_al_2009} The unusual state in S/F/S~JJs is caused by the action of an exchange field~$\mathbf{h}$ in~F on spins of Cooper pairs penetrating into the ferromagnet~F from the singlet superconductors due to the proximity effect. This action leads to spatial oscillations of the wave function of Cooper pairs~$f$ and consequently to the sign change of~$I_{\text{c}}$. Combination of $\pi$- and $0$\nobreakdash-Josephson contacts allows one to construct a so-called $\phi$\nobreakdash-contact, that is, the junctions with a finite arbitrary phase difference~$\phi$ in the ground state.\cite{Buzdin08,Sickinger_et_al_2012,*Goldobin13,Moor13,*Moor_Volkov_Efetov_SUST} Such JJs have a high potential for applications, for example, in realization of the so-called Q\nobreakdash-bits.\cite{SchoenQ}

Another interesting effect occurs in multilayered S/F~JJs if the magnetization vectors~$\mathbf{M}$ in different F~layers are not collinear or the magnetization in the ferromagnet~F is not uniform (helical ferromagnet or ferromagnet with a domain wall).\cite{Bergeret_Volkov_Efetov_2001,BVErmp,Eschrig_Ph_Today} In this case, triplet Cooper pairs arise in the S/F~system with the total spin~$\mathbf{S}$ parallel to~$\mathbf{M}$ in an F~layer which is almost ``transparent'' for these pairs. Even if this ferromagnet is strong, the penetration depth can reach a large value of the order ${\xi \simeq \sqrt{D/T}}$ (in diffusive case) in contrast to a short penetration length ${\xi_{h} \simeq \sqrt{D/h}}$ for singlet Cooper pairs or for the triplet ones with the total spin~$\mathbf{S}$ perpendicular to the vectors~$\mathbf{M}$ (only such Cooper pairs arise in the case of uniform magnetization). The triplet Cooper pairs with ${\mathbf{S} \parallel \mathbf{M}}$ can be called the long-range triplet component (LRTC). The Josephson effect caused by the LRTC has been observed in many experiments on S/F~JJs with nonhomogeneous~$\mathbf{M}$ in ferromagnetic layer(s)\cite{Keizer06,Aarts10,Birge10,Birge12,Zabel10,Petrashov06,Halasz_et_al_2011,Petrashov11,Blamire10,Leksin_et_al_2012} and extensively studied in theoretical works, see, e.g., Refs.~\onlinecite{Eschrig03,Braude07,Tanaka07,Buzdin07,Zaikin08,Valls12,Beri_et_al_2009,Brouwer11} and many other papers cited in reviews Refs.~\onlinecite{BVErmp,Linder_Robinson_2015}, and especially in Ref.~\onlinecite{Eschrig_Reports_2015}. Spin non-dissipative current also arises in such JJs, and therefore these structures with the triplet spin-polarised component may be used in superconducting spintronics.\cite{Linder_Robinson_2015,Eschrig_Reports_2015} A special attention is paid nowadays to the study of spin-orbit interaction in S/F~structures which is necessary for achieving so-called Majorana states.\cite{Galitski12,Bergeret13,Liu14}

Many works are related with strong efforts to detect these exotic quasiparticles---the so-called Majorana fremions---in condensed matter with the help of Josephson junctions as the latter represent a sensitive and convenient tool to achieve this goal. These particles, which are identical to their antiparticles, were predicted long ago,\cite{Majorana37} but only relatively recently it has been shown that they can exist in condensed matter.\cite{SarmaRMP08} In particular, the Josephson coupling may be realized through the Majorana fermions leading to the so-called fractional Josephson effect, i.e., the Josephson current~$I_{\text{J}}$ is related to the phase difference~$\varphi$ as (see Refs.~\onlinecite{Fu_Kane_2008,Tanaka_Yokoyama_Nagaosa_2009,Lutchyn_Sau_DSarma_2010,Oreg_Gil_Oppen_2010})
\begin{equation}
I_{\text{J}}^{\text{M}} = I_{\text{c}} \sin (\varphi /2) \,, \label{0}
\end{equation}
in contrast to the ordinary Josephson current-phase relation\cite{Josephson_1962_original}
\begin{equation}
I_{\text{J}} = I_{\text{c}} \sin (\varphi) \,. \label{01}
\end{equation}
Although some indications on the existence of the Majorana fermions have been obtained in experiments,\cite{MajoranaExp,Anindya_et_al_2012,Deng_et_al_2012,Finck_et_al_2013,Churchill_et_al_2013} further work is needed to make decisive conclusions.

In principle, the current-phase relation can be more complicated in different types of JJs and contain many higher harmonics, ${I_{\text{J}} \propto \sum_{n \geq 0} I_{\text{n}} \sin [ (2n + 1) \varphi]}$.\cite{Likharev,Barone_Paterno_2005}

Interesting physics and new possibilities for application occur in multi-terminal S/n~or S/F~structures. Additional terminals in the JJs allow one to control and to tune the critical Josephson current. For example, a sign-reversal of the Josephson critical current has been observed in multi-terminal S/n/S~JJs with two additional lateral normal terminals\cite{KlapNature99} when voltage was applied to these normal terminals. Theoretically, this nonequilibrium effect has been predicted and studied in Refs.~\onlinecite{V95,Yip98,Wilhelm98}. An inverse effect---a modulation of the conductance between normal reservoirs in the presence of the phase difference between superconductors---has been observed earlier by Petrashov~\emph{et.~al.}\cite{Petrashov95,Petrashov00}

Four-terminal JJs with all superconducting reservoirs have been studied in Refs.~\onlinecite{Amin01,Amin02,Linder12,Mai13,Nazarov15}. The Josephson coupling between different superconductors has been provided through the n~or F~wires connecting the reservoirs. It has been shown that the critical current~$I_{\text{c}}$ may be tuned by varying the phase difference between lateral superconducting reservoirs. The case of two short F~wires connecting the superconductors has been studied in Ref.~\onlinecite{Mai13}. The length of the F~wires has been supposed to be shorter than~$\xi_{h}$ so that the singlet component penetrated into the F~wires due to the proximity effect. Since the magnetisations in crossed wires have been assumed to be perpendicular to each other, not only singlet component but also the LRTC existed in the ferromagnetic wires. The Josephson effect arose due to a complicated interaction between the LRTC with spin-up and spin-down Cooper pairs and the singlet one.

In our recent works,\cite{Moor_Volkov_Efetov_2015_c,Moor_Volkov_Efetov_2015_d} we studied the Josephson effect in two- and three-terminal~S$_{\text{m}}$/I$_{\text{m}}$/n Josephson contacts in which superconducting reservoirs~S$_{\text{m}}$ represent ``magnetic'' superconductors separated from the normal n~wire by a spin filter~I$_{\text{m}}$. [In experiment, the spin filter may be realized by a magnetic insulator or half metal which lets to pass triplet Cooper pairs with only a certain orientation of the total spin~$\mathbf{S}$.] Investigating the Josephson junction of the types S$_{\text{m}}$/I$_{\text{m}}$/n/I$_{\text{m}}$/S$_{\text{m}}$ and S$_{\text{m}}$/F/n/F/S$_{\text{m}}$ (where F represents a strong ferromagnet) we showed that there is a great difference between them.\cite{Moor_Volkov_Efetov_2015_c} The second type of JJs can be called nematic as the strong ferromagnet~F passes the triplet Cooper pairs with spin-up and spin-down orientation (the filter axes are oriented along the $z$\nobreakdash-axis), while the first one is denoted as ``magnetic'' type since the direction of the vector~$\mathbf{S}$ is determined by the orientation of the filter axes~$\mathbf{h}$. In particular, if the $\mathbf{h}$~vectors are antiparallel to each other, there is no Josephson coupling between the right and left superconductors~S$_{\text{m}}$ and the Josephson current is zero, ${I_{\text{J}} = 0}$. If an additional terminal in the form of a singlet superconductor is attached to the n~wire in the S$_{\text{m}}$/I$_{\text{m}}$/n/I$_{\text{m}}$/S$_{\text{m}}$ contact, the Josephson current can flow between the~S and the~S$_{\text{m}}$ reservoirs, while in the absence of any of the three terminals there is no Josephson current. In this case, the phase relation of the Josephson current is rather unusual,\cite{Moor_Volkov_Efetov_2015_d}
\begin{equation}
I_{\text{J}} = I_{\text{c}} \sin (2 \varphi) \,, \label{02}
\end{equation}
where ${\varphi = (\chi_{\text{R}} + \chi_{\text{L}}) / 2 - \chi_{\text{S}}}$, and $\chi_{\text{R(L)}}$, $\chi_{\text{S}}$ are the phases of the right (left) ``triplet'' S$_{\text{m}}$~superconductors and the singlet superconductor, respectively.

One visualizes the case of the Josephson coupling via Majorana fermions as ``fusion'' of a pair of Majorana fermions.  In case of two S$_{\text{m}}$~superconductors and one singlet superconductor, we have a transformation of two singlet Cooper pairs into two triplet Cooper pairs with antiparallel total spins~$\mathbf{S}$ (a supersinglet) which are transferred to the right~(left) superconductor~S$_{\text{m}}$. Thus, different number of Cooper pairs participate in the Josephson coupling---one half in the Majorana case, Eq.~(\ref{0}), one Cooper pair in the conventional Josephson effect, Eq.~(\ref{01}), or two Cooper pairs in the case considered in Ref.~\onlinecite{Moor_Volkov_Efetov_2015_d} with the current-phase relation Eq.~(\ref{02}).

The Josephson effect in the latter case can be seen as an extension of the family of $n$\nobreakdash-fermion condensate caused Josephson effect with according adaptation of the phase dependence with the sequence given by Eqs.~(\ref{0}),~(\ref{01})~and~(\ref{02}), i.e., the phase dependence is represented by $\sin(n \varphi/2)$, respectively.

\begin{figure}[tbp]
\includegraphics[width=1.0\columnwidth]{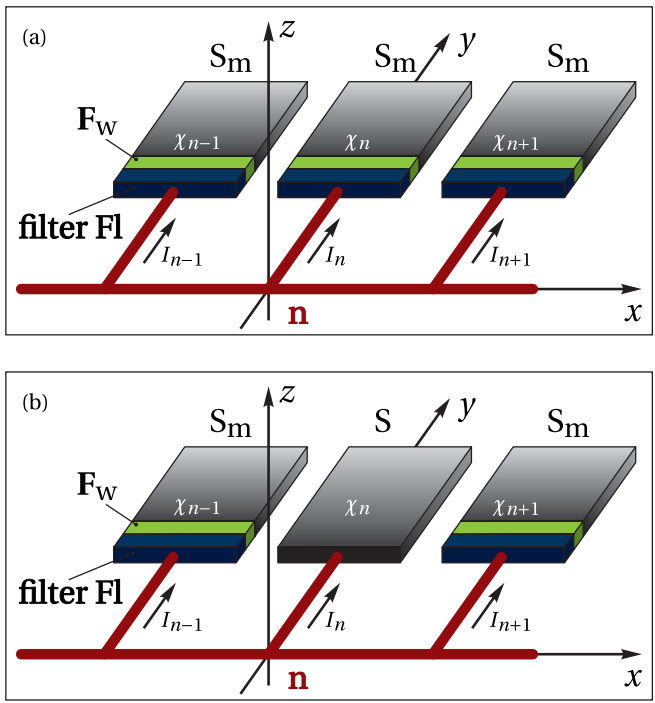}
\caption{(Color online.) Schematic representation of the system under consideration. The rectangles mean the superconductors~S$_{\text{m}}$ or~S constituting the junction with corresponding phases and currents flowing. (a)~Generic multi-terminal Josephson junction consisting of only S$_{\text{m}}$~superconductors; (b)~generic multi-terminal Josephson junction consisting of S$_{\text{m}}$~superconductors plus a singlet superconductor~S.}
\label{fig:System1a}
\end{figure}

In this Paper, we consider generic multi-terminal Josephson junctions consisting of only S$_{\text{m}}$~superconductors [see Fig.~\ref{fig:System1a}~(a)], or of S$_{\text{m}}$~superconductors plus a singlet superconductor [see Fig.~\ref{fig:System1a}~(b)]. The first system allows one to study interaction of fully polarized triplet components, while in the second one we can obtain both---the nematic case with the conventional Josephson relation~$I_{\text{Q}}(\varphi)$, Eq.~(\ref{01}), and also the magnetic case with unusual current-phase relation~$I_{\text{Q}}(\varphi)$, Eq.~(\ref{02}). We show, in particular, that the triplet components with opposite spin direction created by the left and right superconductors~S$_{\text{mL}}$ and~S$_{\text{mR}}$ do not interfere in S$_{\text{mL}}$/n/S$_{\text{mR}}$ or in S$_{\text{mL}}$/F/S$_{\text{mR}}$ JJs at any interface transparencies.

The plan of the paper is as follows. In Section~\ref{sec:model}, we describe the system under consideration and present basic equations. In Section~\ref{sec:two-terminals}, the simplest two-terminal system of the S$_{\text{m}}$/n/S$_{\text{m}}$ or S$_{\text{m}}$/I/S$_{\text{m}}$ type will be studied, where~S$_{\text{m}}$ is a ``magnetic'' superconductor separated from the normal wire~n by a spin-filter. The S$_{\text{m}}$/I/S$_{\text{m}}$~contact is a tunnel junction with a thin insulating layer~I. A general formula for the Josephson current~$I_{\text{Q}}$ in a multi-terminal Josephson junction in the limit of a high S$_{\text{m}}$/n interface resistance will be presented in Section~\ref{sec:multi-terminal}. We use this formula in Section~\ref{sec:four-terminal_triplet-only} to briefly describe the Josephson effect in a four-terminal contact consisting only of S$_{\text{m}}$~superconductors (that is, only fully polarized triplet components exist in the n~wire). The more interesting case of a four-terminal junction with one singlet superconductor and three~S$_{\text{m}}$ superconductors having different orientations of the $\mathbf{h}$~vectors is considered in Section~\ref{sec:four-terminal_with singlet} where the expression for the current is obtained based on the derivation of the general expressions. In Conclusions, we summarize and discuss the obtained results.

\section{Model and basic equations}
\label{sec:model}

We consider a multi-terminal Josephson junction (JJ) which consists of ``magnetic'' superconductors with or without one conventional singlet BCS superconductor (see Fig.~\ref{fig:System1a}). All superconductors are connected by a normal n~wire or film. The ``magnetic'' superconductors are formed by a conventional superconductor covered by a thin ferromagnetic layer~F with an exchange field~$\mathbf{h}$. Due to the proximity effect the singlet component penetrates from the superconductor into the F~film, and also a triplet component arises under the action of the exchange field~$\mathbf{h}$. As is well known (see, e.g., reviews Refs.~\onlinecite{BVErmp,Eschrig_Ph_Today,Eschrig_Reports_2015}), in the case of homogeneous magnetization~$\mathbf{M}$ (with ${\mathbf{M} \parallel \mathbf{h}}$), the vector of the total spin of triplet Cooper pairs~$\mathbf{S}$ lies in the plane perpendicular to~$\mathbf{M}$. Thus, in case of a good contact between the~S and~F layers, the S/F bilayer can be considered as a ``magnetic'' superconductor with a built-in exchange field~$\mathbf{h}$ that has the amplitude ${h_{\text{eff}} = |\mathbf{h}| d_{\text{F}} / (d_{\text{F}} + d_{\text{S}})}$ and a nonzero projection onto the $z$~axis, where~$d_{\text{F(S)}}$ are the thicknesses of the~F and~S layers, respectively.\cite{Bergeret_Volkov_Efetov_2001_b} The F~layer is separated from the n~wire (or film) by a filter that passes electrons only with a certain spin direction, say, parallel or antiparallel to the $z$~axis (filter axis). As a filter, one can use thin layers of strongly polarized magnetic insulator, for example EuO\cite{Moodera08} and DyN or GdN films.\cite{Muduli_et_al_arXiv_2015}

The convenient tool to describe the system under consideration is the method of quasiclassical Green's functions.\cite{RammerSmith,LO,BelzigRev,Kopnin}
This technique has been widely used for studying mesoscopic multi-terminal S/n~structures.\cite{Volkov199321,ZAITSEV1994274,Nazarov_1994,Hekking_Hekking_1993,Hekking_Hekking_1994,Nazarov19991221} In the considered diffusive case, these functions obey the Usadel equation,\cite{Usadel} which in the n~wire has the form
\begin{equation}
- \nabla (\hat{g} \nabla \hat{g}) + \frac{1}{2} \kappa_{\omega }^{2} [ \hat{X}_{30} \,, \hat{g} ] = 0 \,, \label{1}
\end{equation}%
where~${\kappa_{\omega}^{2} = \omega / D}$ with the diffusion coefficient~$D$, and ${\omega = (2n+1) \pi T}$ is the Matsubara frequency. In the considered case of the exchange field acting on the spins of electrons, the Green's function~$\hat{g}$ is a ${4 \times 4}$~matrix in the particle-hole and spin spaces. The matrix ${\hat{X}_{ij} = \hat{\tau}_{i} \cdot \hat{\sigma}_{j}}$ is a tensor product of the Pauli matrices~$\hat{\tau}_{i}$ and~$\hat{\sigma}_{j}$ (${i,j = 0,1,2,3}$) which operate correspondingly in the particle-hole and spin space, respectively, and the $0$\nobreakdash-th Pauli matrix is just the unity ${2 \times 2}$~matrix. Moreover, the matrix quasiclassical Green's function~$\hat{g}$ obeys the normalization condition
\begin{equation}
\hat{g} \cdot \hat{g} = 1 \,. \label{2}
\end{equation}

As in our previous works,\cite{Moor_Volkov_Efetov_2015_c,Moor_Volkov_Efetov_2015_d} we use a representation for the matrix Green's functions~$\hat{g}$ suggested by Ivanov and Fominov.\cite{IvanovFomin} These Green's functions are related to those in Refs.~\onlinecite{Bergeret_Volkov_Efetov_2001,BVErmp}, $\hat{g}_{\text{BVE}}$, via the transformation ${\hat{g} = U \cdot
\hat{g}_{\text{BVE}} \cdot U^{\dagger}}$ with ${U = (1/2) (1 + i \hat{X}_{33}) \cdot (1 - i \hat{X}_{03})}$.

Equation~(\ref{1}) is complemented by boundary conditions at the interfaces S$_{\text{m}}$/n and S/n. They have the form [see Refs.~ \onlinecite{EschrigBC13,*EschrigBC13a,EschrigBC15}, as well as Eq.~(4.7) in Ref.~\onlinecite{Bergeret12b}]
\begin{equation}
    L_{\nu} \hat{g} \partial_{\nu} \hat{g}_{|_{\nu = \pm L_{\nu}}} = \pm r_{\nu} [\hat{g} \,, \hat{\mathrm{\Gamma}}_{\nu} \hat{G}_{\nu} \hat{\mathrm{\Gamma}}_{\nu}]_{|_{\nu = \pm L_{\nu}}} \,,
    \label{3}
\end{equation}
where ${r_{\nu} = L_{\nu}/\sqrt{2}\sigma R_{\text{b},\nu}}$, with the conductivity of the n~wire~$\sigma$ and the n\nobreakdash-S$_{\text{m}}$ interface resistance at~$L_{\nu}$ per unit area~$R_{\text{b},\nu}$. The matrix coefficient~$\hat{\mathrm{\Gamma}}$ describes the electron transmission with a spin-dependent probability~$\mathcal{T}_{\uparrow,\downarrow}$. If the filters let to pass only electrons with spins aligned parallel to the $z$~axis, then ${\hat{\mathrm{\Gamma}} = \mathcal{T} \hat{1} + \mathcal{U} \hat{X}_{33}}$ so that the probability for an electron with spin up (down) to penetrate into the n~wire is ${\mathcal{T}_{\uparrow,\downarrow} \propto \mathcal{T} \pm \mathcal{U}}$. We assume that ${\mathcal{U} = \zeta \mathcal{T}}$ with ${\zeta = \pm 1}$, and that the coefficients~$\mathcal{T}$ and $\mathcal{U}$ are normalized, ${\mathcal{T} = |\mathcal{U}| = \sqrt{2}}$. The S/n interface between the conventional superconductor and the normal metal were is assumed to be spin independent so that ${\hat{\mathrm{\Gamma}} = \hat{1}}$ and $r_{\text{S}} = L/\sigma R_{\text{S}}$.

We have to find a solution of Eq.~(\ref{1}) taking into account the normalization and boundary conditions Eqs.~(\ref{2}) and~(\ref{3}). This can be easily done in the case of a short normal wire, i.e., when the condition ${L \ll \sqrt{D/T}}$ holds. Then, integrating Eq.~(\ref{1}) over the coordinate along the normal wire with account for the boundary conditions yields the equation
\begin{equation}
\lbrack \hat{\Lambda} \,, \hat{g}] = 0 \,, \label{4}
\end{equation}
where the matrix ${\hat{\Lambda} = \hat{\Lambda}_{\text{n}} + \hat{\Lambda}_{\text{m}}}$ is a sum of contributions of the normal wire and S$_{\text{m}}$~superconductors, ${\hat{\Lambda}_{\text{n}} = r_{\omega} \hat{X}_{30}}$ with ${r_{\omega} = \omega L^{2} / D}$. The matrix
\begin{equation}
\hat{\Lambda}_{\text{m}} = \sum_{\nu} \hat{\Lambda}_{\nu} \label{4a}
\end{equation}
is related to the Green's functions in the $\nu$\nobreakdash-th ``magnetic'' superconductor~$\hat{G}_{\text{m}}$ via ${\hat{\Lambda}_{\nu} = r_{\nu} [\hat{\Gamma} \cdot \hat{G}_{\text{m}} \cdot \hat{\Gamma}]_{\nu}}$, or in the singlet superconductor via ${\hat{\Lambda}_{\text{S}} = r_{\text{S}} \hat{G}_{\text{S}}}$. The Green's function ${\mathbf{\hat{G}}_{\nu} \equiv [ \hat{\Gamma} \cdot \hat{G}_{\text{m}} \cdot \hat{\Gamma}]_{\nu}}$ in the S$_{\text{m}}$~superconductors have the form\cite{Moor_Volkov_Efetov_2015_c,Moor_Volkov_Efetov_2015_d}
\begin{equation}
\mathbf{\hat{G}}_{\nu} = g_{+} \hat{X}_{30} + g_{-} \hat{X}_{33} + \hat{F}_{\nu} \,, \label{5}
\end{equation}
where the condensate Green's function in the ``magnetic'' superconductor~$\hat{F}_{\nu}$ with the phase~$\chi_{\nu}$ is defined as
\begin{equation}
\hat{F}_{\nu} = f_{-} \exp [i \chi_{\nu} \hat{X}_{30}] \cdot \hat{X}_{\nu} \,,  \label{5'}
\end{equation}
where
\begin{align}
f_{\pm} &= [f(\omega + ih) \pm f(\omega - ih)]/2 \,, \label{eq:f_pm} \\
g_{\pm} &= [g(\omega + ih) \pm g(\omega - ih)]/2 \,, \label{eq:g_pm}
\end{align}
with ${f(\omega) = (\Delta / \omega) \cdot g(\omega) = \Delta / \sqrt{\omega^{2} + \Delta^{2}}}$. The form of the matrices~$\hat{X}_{\nu}$ depends on the chirality of the triplet component\cite{Moor_Volkov_Efetov_2015_d} and on the direction of the filter axes, i.e., on the sign of~$\zeta$. In the case of the $x$- or $y$\nobreakdash-chirality (i.e., the vector~$\mathbf{h}$ is directed along the $x$- or $y$\nobreakdash-axis, respectively) this matrices have the form
\begin{align}
\hat{X}_{x}(\zeta) &= \hat{X}_{11} - \zeta \hat{X}_{22} \,,
\label{6} \\
\hat{X}_{y}(\zeta) &= \hat{X}_{12} + \zeta \hat{X}_{21} \,.
\label{6'}
\end{align}
In the case of the $x$\nobreakdash-chirality, the filter lets to pass Cooper pairs with spin up if ${\zeta = +1}$ and the S$_{\text{m}}$/n interface is transparent only for the Cooper pairs with spin down if ${\zeta = -1}$, and vice versa for the $y$\nobreakdash-chirality. One can show that terms given by~$\hat{X}_{x}(+1)$ describe correlators of the form ${\propto \langle \hat{c}_{\uparrow} \hat{c}_{\uparrow}(t)\rangle}$ while those given by~$\hat{X}_{x}(-1)$---correlators of the form ${\propto \langle \hat{c}_{\downarrow} \hat{c}_{\downarrow}(t)\rangle}$; correspondingly, describe the terms given by~$\hat{X}_{y}(-1)$ correlators of the form ${\propto \langle \hat{c}_{\uparrow} \hat{c}_{\uparrow}(t)\rangle}$ while those given by~$\hat{X}_{y}(+1)$---correlators of the form ${\propto \langle \hat{c}_{\downarrow} \hat{c}_{\downarrow}(t)\rangle}$.

Knowing the Green's function $\hat{g} \equiv \hat{g}_{\text{d}} + \hat{f}$ in the n~wire (here,~$\hat{g}_{\text{d}}$ and~$\hat{f}$ are diagonal and, respectively, off-diagonal in the Gor'kov-Nambu space parts of~$\hat{g}$), one can easily find the Josephson charge~$I_{\text{Q}}$ current at the interface of the $\nu$\nobreakdash-th terminal, which is given by the expression
\begin{equation}
I_{\text{Q}|_{\nu}} = i a r_{\nu} (2 \pi T) \sum_{\omega} \mathrm{Tr} \{ \hat{X}_{30} \cdot [ \hat{f} \,, \hat{F}_{\nu} ] \} \,, \label{I_Q} \\
\end{equation}
where ${a = \sigma/(16 e L)}$, and~$\sigma$ is the conductivity of the n~wire.

\begin{figure}
  \centering
  \includegraphics[width=1.0\columnwidth]{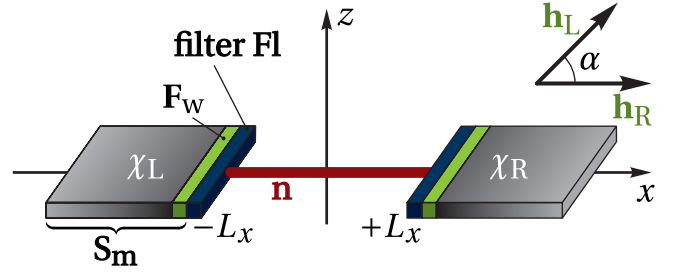}
  \caption{(Color online.) Schematic representation of a simple S$_{\text{m}}$/n/S$_{\text{m}}$ Josephson junction. The filters~Fl, denoted by blue layers attached to the n~wire, can be oriented parallel (${\zeta_{\text{R}} = \zeta_{\text{L}}}$) or antiparallel (${\zeta_{\text{R}} = - \zeta_{\text{L}}}$).}
  \label{fig:setup4}
\end{figure}

To begin with, we consider the simplest case of a two-terminal Josephson junction, i.e., the S$_{\text{m}}$/n/S$_{\text{m}}$ contact schematically sketched in Fig.~\ref{fig:setup4}.

\section{Two-terminal Josephson contact}
\label{sec:two-terminals}

\subsection{Josephson current in S$_{\text{m}}$/Fl/n/Fl/S$_{\text{m}}$ junctions}

The dc Josephson effect in this system has been considered in our previous work\cite{Moor_Volkov_Efetov_2015_c} under assumption of a small transparency of the right (left) S$_{\text{m}}$/n interfaces, ${r_{\text{R}} = r_{\text{L}} \equiv r_{\text{m}} \ll 1}$). There, it has been shown that the Josephson current is zero in case of antiparallel (${\zeta_{\text{R}} = - \zeta_{\text{L}}}$) and is finite in JJs with parallel ($\zeta_{\text{R}} = \zeta_{\text{L}}$) filter axes orientations. Here, we show that in case of a short normal wire or a thin n~film (${L \ll \xi_{T} = \sqrt{D / \pi T}}$), this statement remains valid for arbitrary transmittance coefficient~$r_{\text{m}}$. In addition to the Josephson current, we calculate also the density of states (DOS) in the n~wire for parallel (${\zeta_{\text{R}} = \zeta_{\text{L}}}$) and antiparallel (${\zeta_{\text{R}} = - \zeta_{\text{L}}}$) filter axes orientations.

We need to solve Eq.~(\ref{4}), where the matrix ${\hat{\Lambda} = \hat{\Lambda}_{\text{n}} + \hat{\Lambda}_{\text{m}}}$ is a sum of two matrices,~$\hat{\Lambda}_{\text{n}}$ and ${\hat{\Lambda}_{\text{m}} = \hat{\Lambda}_{\text{L}} + \hat{\Lambda}_{\text{R}}}$. The matrices~$\hat{\Lambda}_{\text{R(L)}}$ at the right~(left) S$_{\text{m}}$/n interfaces are related to the Green's functions~$\hat{G}_{\text{m}}$ in ``magnetic'' superconductors as
\begin{align}
\hat{\Lambda}_{\text{n}} &= 2 r_{\omega} \hat{X}_{30} \,, \label{7} \\
\hat{\Lambda}_{\text{R(L)}} &= r_{\text{m}} [ \hat{\Gamma} \cdot \hat{\mathbf{G}}_{\text{m}} \cdot \hat{\Gamma} ]_{\text{R(L)}} \,.  \label{7'}
\end{align}

The Green's function~${\hat{\mathbf{G}}_{\text{m}} \equiv \hat{R}_{\pi/2, 2} \cdot \hat{G}_{\text{m}} \cdot \hat{R}_{\pi /2, 2}^{\dagger} }$ in the S$_{\text{m}} $ superconductor is related to the Green's function~$\hat{G}_{\text{m}}$ in a superconductor with a uniform exchange field~$h$ oriented along the $z$\nobreakdash-axis by means of the rotation matrix ${\hat{R}_{\pi /2, 2} = \cos (\pi /4) + i\hat{X}_{02} \sin (\pi /4)}$. The form of the rotation matrix means that we assume for definiteness the $x$\nobreakdash-chirality for the triplet component in both S$_{\text{m}}$~superconductors.\cite{Moor_Volkov_Efetov_2015_d} Thus, we find for~$\hat{\Lambda}_{\text{R(L)}}$
\begin{equation}
\hat{\Lambda}_{\text{R(L)}} = r_{\text{R(L)}} [g_{+} ( \hat{X}_{30} + \zeta_{\text{R(L)}} \hat{X}_{03} ) + f_{-} \exp (i \chi_{\text{R(L)}} \hat{X}_{30} ) \cdot \hat{X}_{\text{R(L)}} ] \,,  \label{7a}
\end{equation}
where~$\hat{X}_{\text{R(L)}}$ is one of the matrices defined in Eqs.~(\ref{6}) and~(\ref{6'}), depending on the chirality of the triplet components generated by the the right, respectively, left S$_{\text{m}}$~superconductor. The final results do not depend on the type of chiralitiy and, thus, we assume them to be equal.

Knowing the matrix~$\hat{g}$, we can find the density of states~$\nu(\varepsilon)$ in the normal wire wire,
\begin{equation}
\nu(\varepsilon) = \frac{\Re\big( \mathrm{Tr} \{ \hat{X}_{30} \hat{g} \}_{|_{\omega = -i \varepsilon}} \big) }{4} \,, \label{DOS1}
\end{equation}
and the Josephson current~$I_{\text{Q}}$ in the system,
\begin{align}
I_{\text{Q}} &= i a (1/4) (2 \pi T) \sum_{\omega = 0} \mathrm{Tr} \{ \hat{X}_{30} [\hat{g} \,, \hat{g}_{\text{R}}] \} \label{I_Q1} \\
&= i a (1/4) (2 \pi T) \sum_{\omega = 0} \mathrm{Tr} \{\hat{X}_{30} [\hat{f} \,, \hat{f}_{\text{R}} ] \} \,, \notag
\end{align}
where ${\hat{g}_{\text{R(L)}} = \hat{g}(\pm L_{x})}$ are the Green's functions at the right (left) S$_{\text{m}}$/n interface while~$\hat{f}$ and~$\hat{f}_{\text{R}}$ are the condensate (off-diagonal in the particle-hole space) parts of the Green's functions in the n~wire and in the right~S$_{\text{m}}$, respectively.

Equation~(\ref{4}) can be solved for an S$_{\text{m}}$/n/S$_{\text{m}}$ Josephson junction in a general case, but we present here simple analytical results for some particular cases. The details of the derivation in each case are provided in the Appendix~\ref{app:technical_details}.

\paragraph{Antiparallel filter orientation (${\zeta_{\text{R}} = - \zeta_{\text{L}} \equiv \zeta}$).}

\begin{figure}
  \centering
  \includegraphics[width=1.0\columnwidth]{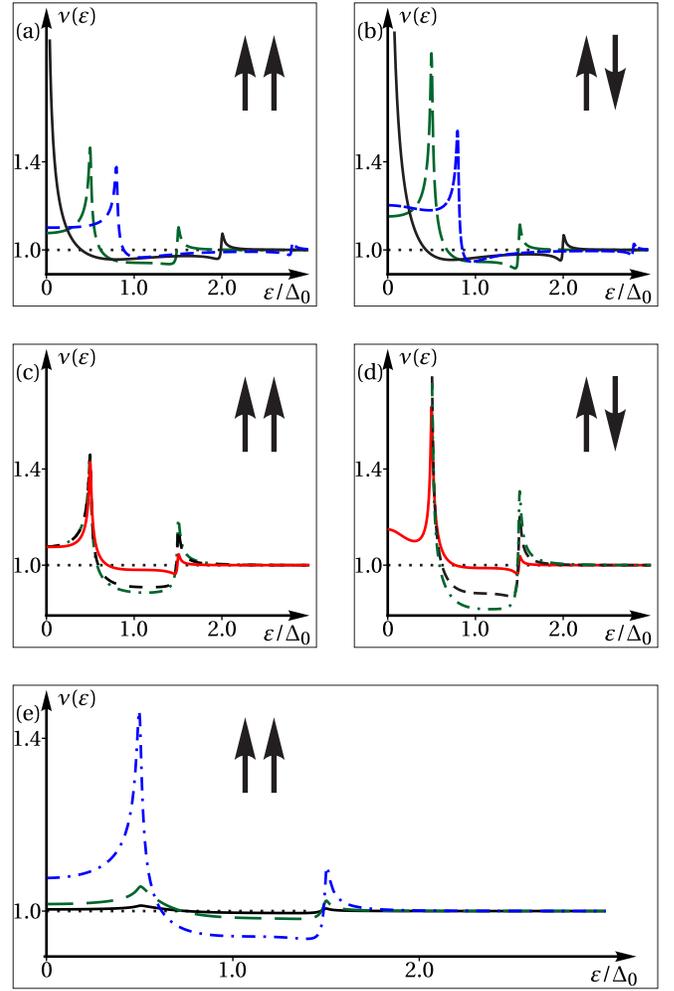}
  \caption{(Color online.) Density of states in the two-terminal Josephson contact made of S$_{\text{m}}$~superconductors. The arrows indicate the filter configuration, i.e., parallel [panels~(a), (c), and~(e), from~Eq.~(\ref{16})] and antiparallel [panels~(b) and~(d), from~Eq.~(\ref{9})]. The top row displays the DOS for different values of ${h = |\mathbf{h}|}$, i.e., ${h = 0}$ (black dotted line); ${h = 0.5 \Delta_0}$ (green long-dashed line); ${h = 1.0 \Delta_0}$ (black solid line); ${h = 1.8 \Delta_0}$ (blue short-dashed line). Additional parameters are ${\cos \varphi = 1.0}$ and ${r_{\text{m}} = 0.25}$. In both the cases the peaks are located at ${\varepsilon = |\Delta_0 \pm h|}$. The middle row shows the DOS for different values of~$r_{\text{m}}$, i.e., ${r_{\text{m}} = 0.1}$ (red solid line); ${r_{\text{m}} = 0.5}$ (black dashed line); ${r_{\text{m}} = 1.0}$ (green dash-dotted line); the black dotted line indicates ${\nu = 1}$. Here, ${\cos \varphi = 1.0}$ and ${h = 0.5 \Delta_0}$. The bottom row shows the DOS for different values of ${\cos \varphi = 1.0}$ in the case of parallel orientation of filters, i.e., ${\cos \varphi = 0}$ (black dotted line); ${\cos \varphi = 0.25}$ (black solid line); ${\cos \varphi = 0.5}$ (green dashed line); ${\cos \varphi = 1.0}$ (blue dash-dotted line). Here, ${h = 0.5 \Delta_0}$ and ${r_{\text{m}} = 0.25}$. In the case of antiparallel filter orientation, there is no dependence of DOS on~$\varphi$.}
  \label{fig:DoS}
\end{figure}

The density of states is calculated from Eqs.~(\ref{DOS1}) and~(\ref{8'}),
\begin{equation}
\nu(\varepsilon) = \Re\big[ (1 + \gamma_{\text{a}}^{2})^{-1/2}_{|_{\omega=-i \varepsilon}} \big] \,, \label{9}
\end{equation}
with~$\gamma_{\text{a}}$ defined in Eq.~(\ref{eq:gamma_a_appendix}). In this case of antiparallel filter orientation, the DOS does not depend on the phase difference~${\varphi = \chi_{\text{R}} - \chi_{\text{L}}}$. In Fig.~\ref{fig:DoS}, we plot the energy dependence of the DOS for both the configurations of the filters.

In the considered system with antiparallel spin filter axes the Josephson current is zero at any transparencies of the S$_{\text{m}}$/n interfaces,
\begin{equation}
I_{\text{Q}} = 0 \,. \label{I_Qa}
\end{equation}

\begin{figure}
  \centering
  \includegraphics[width=1.0\columnwidth]{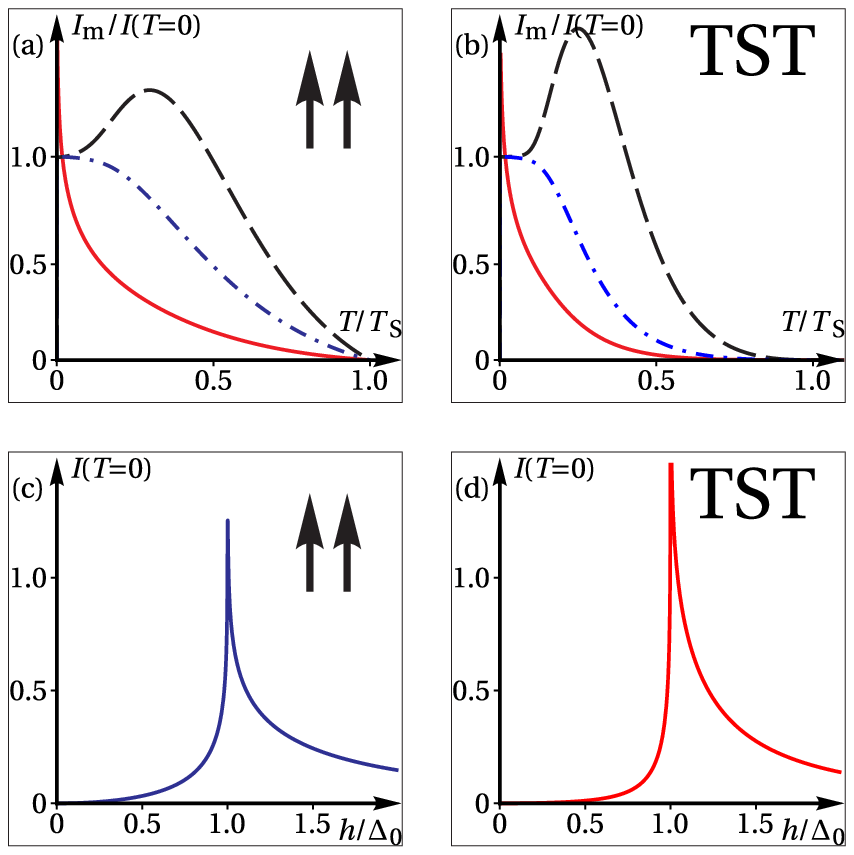}
  \caption{(Color online.) Critical current in the in the two-terminal Josephson contact with parallel filter directions [panels~(a) and~(c), from~Eq.(\ref{I_c_b})] and in the three-terminal contact~TST consisting of two ``magnetic'' S$_{\text{m}}$~superconductors and a singlet superconductor~S [panels~(b) and~(d), from~Ref.~\onlinecite{Moor_Volkov_Efetov_2015_d}]. In the two-terminal Josephson contact with antiparallel filter directions, ${I_{\text{Q}} = 0}$, see Eq.~(\ref{I_Qa}). Top panels show the temperature dependence of~$I_{\text{c}}$ for different values of ${h = |\mathbf{h}|}$, i.e., ${h = 1.5 \Delta_0}$ (blue dash-dotted line); ${h = 1.0 \Delta_0}$ (red solid line); ${h = 0.5 \Delta_0}$ (black dashed line). Noticeably, at ${h < \Delta_0}$, the temperature dependence of~$I_{\text{c}}$ is nonmonotonic and has a maximum at some temperature. The bottom panels display the dependence of~$I_{\text{c}}$ on~$h$ at ${T = 0}$ resembling the Riedel peak.\cite{Riedel64} Other parameters are ${\cos \varphi = 1.0}$ and ${r_{\text{m}} = 0.25}$.}
  \label{fig:Current}
\end{figure}

\paragraph{Parallel filter orientation (${\zeta_{\text{R}} = \zeta_{\text{L}} \equiv \zeta}$).}

The density of states, Eq.~(\ref{DOS1}), in this case is given by
\begin{align}
\nu(\varepsilon) &= \varrho \cos \alpha \label{16} \\
&= \frac{1}{2} \Re \big[ 1 + [1 + 4 \gamma_{\text{b}}^{2}(\varphi) ]^{-1/2}_{|_{\omega =-i \varepsilon}} \big] \notag
\end{align}
with~$\gamma_{\text{b}}$ defined in Eq.~(\ref{13}). The Josephson current is
\begin{equation}
I_{\text{Q}} = I_{\text{c}}(\varphi) \sin \varphi \,, \label{I_Qb}
\end{equation}
with the critical current
\begin{equation}
I_{\text{c}}(\varphi) = a r_{\text{m}}^{2}(2 \pi T) \sum_{\omega \geq 0} \frac{f_{-}^{2}}{\big(\frac{\omega}{\Delta} + 2 r_{\text{m}} \frac{E_{\text{Th}}}{\Delta} g_{+} \big) \sqrt{1 + 4 \gamma_{\text{b}}^{2}(\varphi)}} \,.  \label{I_c_b}
\end{equation}

One can see that the phase dependence of~$I_{\text{c}}$ leads to appearance of higher harmonics in the Josephson current, i.e., the current~$I_{\text{Q}}$ in Eq.~(\ref{I_Qb}) can be written as
\begin{equation}
I_{\text{Q}} = \sum_{n = 1}^{\infty} I_n \sin[(2 n + 1) \varphi] \,. \label{I_Qb_harmonics}
\end{equation}

In Fig.~\ref{fig:DoS}, we plot the energy dependence of the DOS for both the configurations of the filters, i.e., parallel [panels~(a), (c), and~(e)] and antiparallel [panels~(b) and~(d)]. The top row displays the DOS for different values of ${h = |\mathbf{h}|}$. Additional parameters~$\varphi$ and~$r_{\text{m}}$ are fixed. In both the cases the peaks are located at ${\varepsilon = |\Delta_0 \pm h|}$. The middle row shows the DOS for different values of~$r_{\text{m}}$ with fixed~$\varphi$ and~$h$. The bottom row shows the DOS for different values of~$\varphi$ in the case of parallel orientation of filters with fixed~$h$ and~$r_{\text{m}}$. In the case of antiparallel filter orientation, there is no dependence of DOS on~$\varphi$. Observe that the dependence~$\nu(\varepsilon)$ is similar for parallel and antiparallel filter orientations---excluding the dependence of the amplitude of DOS on the phase which is there for the parallel case, but absent in the antiparallel filter configuration.

Thus, in order to experimentally distinguish the both cases, it is safer to rely on measurements of the Josephson critical current displayed in Fig.~\ref{fig:Current} in the left column, [panels~(a) and~(c), from~Eq.(\ref{I_c_b})], for parallel filter configuration. In the case of antiparallel filter orientation, ${I_{\text{Q}} = 0}$, see Eq.~(\ref{I_Qa}). We compare the dependence of the Josephson critical current with the case of a so-called TST-contact with T~denoting the ``magnetic'' superconductor with filters oriented antiparallel and S~is an additional singlet superconducting reservoir attached to the normal wire.\cite{Moor_Volkov_Efetov_2015_d} Top panels show the temperature dependence of~$I_{\text{c}}$ for different values of~$h$ with fixed~$\varphi$ and~$r_{\text{m}}$. Noticeably, at ${h < \Delta_0}$, the temperature dependence of~$I_{\text{c}}$ is nonmonotonic and has a maximum at some temperature. The bottom panels display the dependence of~$I_{\text{c}}$ on~$h$ at ${T = 0}$ resembling the Riedel peak,\cite{Riedel64} but in our case, the role of the voltage~$V$ is played by the exchange field~$h$ in the weak ferromagnet~F$_{\text{w}}$ responsible for creation of the triplet component. Also, the location of the peak is given by ${h = \Delta}$ in contrast to ${V = 2 \Delta}$ as it is the case in the Riedel singularity.\cite{Riedel64}

In order to facilitate the comparison of the critical currents in the considered systems with that in a conventional Josephson junction of the S/n/S type, we now present the DOS and the critical current for such a junction. Again, the details of the derivation are provided in the Appendix~\ref{app:technical_details}.

\paragraph{Usual S/n/S Josephson junction.}

In this case, the density of states is given by the expression
\begin{equation}
\nu(\varepsilon) = \Re\big[ [1 + \gamma_{\text{c}}^{2}(\varphi)]^{-1/2}_{|_{\omega=-i \varepsilon}} \big] \,, \label{19}
\end{equation}
with~$\gamma_{\text{c}}$ given in Eq.~(\ref{eq:gamma_c_appendix}), and the Josephson current reads
\begin{align}
I_{\text{J}} &= I_{\text{c}} \sin \varphi \,, \label{20} \\
\intertext{with}
I_{\text{c}} &= 2 a r_{\text{S}}^{2} (2 \pi T) \sum_{\omega \geq 0} \frac{F_{\text{S}}^{2}}{\sqrt{ 1 + \gamma_{\text{c}}^{2}(\varphi) }} \,. \label{20'}
\end{align}
The coefficient~$a$ is the same as in Eq.~(\ref{I_c_b}) and related with the critical current~$I_{\text{c}}$ via Eq.~(\ref{20}).

As concerns the temperature and $h$~dependence of the critical current~$I_{\text{c}}$ for the parallel filter axes given by Eq.~(\ref{20}), we see that the temperature dependence is not monotonous and has a maximum at a temperature below~$T_{\text{c}}$. Similar dependencies have been obtained for a ballistic Josephson junction with spin-active interfaces.\cite{Eschrig03,Eschrig_Reports_2015} This maximum has been interpreted as a contribution of the Andreev bound states to the Josephson current. In our diffusive case there are no Andreev bound states, so this explanation is not universal. In our system, it is related to a singularity in the DOS and in the Green's functions at ${h = \Delta}$. This singularity resembles the Riedel singularity on the $I$\nobreakdash-$V$~characteristics if a voltage~$V$ is applied to a junction.\cite{Riedel64} To some extent, from the mathematical point of view, the voltage~$V$ is analogous to the exchange field~$h$. The form of the $I$\nobreakdash-$V$~curve for an S/F/S Josephson contact has been found in Ref.~\onlinecite{Bobkov06}.

In order to make the nature of the maximum in the $I_{\text{c}}(T)$~dependence more clear, we present in the next subsection the critical current for different tunnel JJs with ``magnetic'' superconductors with and without spin filters.

\subsection{Josephson current in tunnel S$_{\text{m}}$/Fl/I/Fl/S$_{\text{m}}$ junctions}

As has been found earlier, a singular behavior of the Green's functions in S$_{\text{m}}$/I/S$_{\text{m}}$ junctions with antiparallel orientations of the  $\mathbf{h}$~vectors leads to an enhancement of the critical Josephson current.\cite{Bergeret_Volkov_Efetov_2001_b} The Josephson current in S$_{\text{m}}$/I/S$_{\text{m}}$~JJs is given again by Eq.~(\ref{I_Q1}) with ${\hat{f} = \hat{f}_{\text{L}}}$ and
\begin{equation}
\hat{f}_{\text{L(R)}} = \exp (i \chi_{\text{L(R)}} \hat{X}_{30}) [ f_{+} \hat{X}_{10} \pm f_{-} \hat{X}_{13} ]_{\text{L(R)}} \,. \label{21a}
\end{equation}

Assuming the $\mathbf{h}$~vectors parallel to the $z$~axis we have
\begin{equation}
f_{\pm |_{\text{L(R)}}} = \Delta_{\text{L(R)}} \frac{\big[ \zeta_{+}^{-1} \pm \zeta_{-}^{-1}\big]_{|_{\text{L(R)}}}}{2} \,,
\end{equation}
with ${\zeta_{\pm |_{\text{L(R)}}} = \sqrt{(\omega \pm ih)^{2} + \Delta^{2}}_{|_{\text{L(R)}}}}$. Simple calculations yield for the Josephson current
\begin{align}
I_{\text{J}} &= I_{\text{c}} \sin \varphi \,, \\
\intertext{with the critical current for parallel ($I_{\text{c} \uparrow \uparrow}$) respectively antiparallel ($I_{\text{c} \uparrow \downarrow}$) $\mathbf{h}$~orientations}
I_{\text{c} \uparrow \uparrow} &\propto (2 \pi T) \Delta_{\text{R}} \Delta_{\text{L}} \sum_{\omega \geq 0} \Re \big( \zeta_{+\text{R}} \big) \Re \big( \zeta_{+\text{L}} \big) \big/ D(\omega) \,, \label{22a} \\
I_{\text{c} \uparrow \downarrow} &\propto (2 \pi T) \Delta_{\text{R}} \Delta_{\text{L}} \sum_{\omega \geq 0} \Re \big( \zeta_{+\text{R}} \zeta_{-\text{L}} \big) \big/ D(\omega) \,, \label{22b}
\end{align}
where ${D(\omega) = (\zeta_{+\text{R}} \zeta_{-\text{R}})(\zeta_{+\text{L}} \zeta_{-\text{L}})}$. Both critical currents,~$I_{\text{c} \uparrow \uparrow}$ and~$I_{\text{c} \uparrow \downarrow}$, occur due to tunneling of singlet and triplet components.

We present here also the expression for the critical current in a S$_{\text{m}}$/Fl/I/Fl/S$_{\text{m}}$ contact, where~S$_{\text{m}}$/Fl is a ``magnetic'' superconductor~S$_{\text{m}}$ with a spin filter~Fl. Then, the Josephson current us caused only by tunneling of triplet Cooper pairs. In the case of filters passing only triplet Cooper pairs with the total spin parallel to the $x$~axis ${\hat{f}_{\text{L(R)}} = f_{- \text{L(R)}} \exp (i \chi_{\text{L(R)}} \hat{X}_{30}) \cdot (\hat{X}_{11} - \hat{X}_{22})}$ and the critical current is
\begin{equation}
I_{\text{c},\text{m} \uparrow \uparrow} \propto - (2 \pi T) \sum_{\omega \geq 0} \Im \big( \zeta_{+\text{R}} \big) \Im \big( \zeta_{+\text{L}} \big) \big/ D(\omega) \,. \label{23}
\end{equation}
In case of antiparallel filter axes, we again have
\begin{equation}
I_{\text{c},\text{m} \uparrow \downarrow} = 0 \,. \label{23_antiparallel}
\end{equation}

The critical current~$I_{\text{c} \uparrow \uparrow}(h)$ has no peak as a function of~$h$, while at low temperatures the critical current~$I_{\text{c} \uparrow \downarrow}(h)$, Eq.~(\ref{22b}), has a sharp peak similar to the one shown in Fig.~\ref{fig:Current}~(c). Both currents,~$I_{\text{c} \uparrow \uparrow}$ and~$I_{\text{c} \uparrow \downarrow}$ decay monotonously with increasing temperature.

The critical current~$I_{\text{c},\text{m} \uparrow \uparrow}$ in Eq.~(\ref{23}) in an S$_{\text{m}}$/Fl/I/Fl/S$_{\text{m}}$ contact with parallel filter axes has a peak as a function of~$h$ and, contrary to~$I_{\text{c} \uparrow \uparrow}(h)$ and~~$I_{\text{c} \uparrow \downarrow}(h)$, has a maximum in the temperature dependence. This behavior is similar to the case illustrated in Figs.~\ref{fig:Current}~(a)~and~\ref{fig:Current}~(c) for the critical Josephson current~$I_{\text{c}}$ in an S$_{\text{m}}$/Fl/n/Fl/S$_{\text{m}}$ contact, see Eq.~(\ref{I_c_b}).

\section{Charge currents in multi-terminal systems}
\label{sec:multi-terminal}

In this section, we consider a multi-terminal Josephson junction of a type shown in Fig.~\ref{fig:System1a} and calculate the charge currents through $n$~different S$_{\text{m}}$/n~interfaces assuming coefficients~$r_{n}$ as small parameters. We consider a system with many~S$_{\text{m}}$ superconductors and with one singlet superconductor~S, which is coupled to the n~wire via the coefficient~$r_{\text{S}}$. The transmittance of the S/n~interface can be arbitrary, i.e., the coefficient~$r_{\text{S}}$ varies from~$0$ (no coupling) to~$\infty $ (perfect S/n interface). In this case, the solution of Eq.~(\ref{4}) is given by
\begin{equation}
\delta \hat{f} = \frac{1}{2 \mathcal{E}} \big[ \hat{\Lambda}_{\text{m}} - \hat{g}_{0} \cdot \hat{\Lambda}_{\text{m}} \cdot \hat{g}_{0} \big] \,, \label{24}
\end{equation}
where ${\mathcal{E} = \sqrt{\tilde{G}_{\text{S}}^{2} + F_{\text{S}}^{2}}}$, ${\hat{g}_{0} = \hat{X}_{30} \tilde{G}_{\text{S}} + \exp ( i \chi \hat{X}_{30} ) \cdot \hat{X}_{10} F_{\text{S}}}$, $\chi$~is the phase of the singlet superconductor, and the functions~$\tilde{G}_{\text{S}}$ and~$F_{\text{S}}$ are defined in Eq.~(\ref{17}).

Using Eqs.~(\ref{7a}) and~(\ref{24}), we find the correction to the condensate Green's function in the n~wire due to the presence of ``magnetic'' superconductors S$_{\text{m}}$ given by the matrix~$\hat{\Lambda}_{\text{m}}$,
\begin{equation}
\delta \hat{f} = \sum_{n^{\prime}} \delta \hat{f}_{n^{\prime}} \,, \label{25'}
\end{equation}
with
\begin{widetext}
\begin{equation}
\delta \hat{f}_{n^{\prime}} = \frac{f_{-}}{2 \mathcal{E}^{3}} r_{n^{\prime}} \big[ [A \exp(i \chi_{n^{\prime}} \hat{X}_{30}) \hat{X}_{n^{\prime}}(\zeta_{n^{\prime}})] - B \exp[ i (2 \chi - \chi_{n^{\prime}}) \hat{X}_{30} ] \cdot \hat{X}_{n^{\prime}}(\zeta_{n^{\prime}}) \big] \,, \label{25}
\end{equation}
\end{widetext}
where ${A = 2 \tilde{G}_{\text{S}}^{2} + F_{\text{S}}^{2}}$ and ${B = F_{\text{S}}^{2}}$. The current through the $n$\nobreakdash-th contact is\footnote{The spin current~$I_{\text{Sp}|_{n}}$ can be calculated in an analogous way substituting ${\mathrm{Tr} \big\{ \hat{X}_{30} \big[ \ldots \big] \big\}}$ by ${\mathrm{Tr} \big\{ \hat{X}_{03} \big[ \ldots \big] \big\}}$ in the expression for the charge current. We restrict ourselves with the calculation of the electric current because experimental technique for observing the spin current is yet not as well developed as for the charge current. Calculations of the spin current for some special simpler cases of considered structures have been performed in Refs.~\onlinecite{Moor_Volkov_Efetov_2015_c,Moor_Volkov_Efetov_2015_d}.}
\begin{align}
I_{\text{Q} |_{n}} &= \sum_{n^{\prime}} I_{n n^{\prime}} \,, \\
\intertext{with}
I_{n n^{\prime}} &= i a r_{n} (2 \pi T) \sum_{\omega, n^{\prime}} f_{-} \mathrm{Tr} \big\{ \hat{X}_{30} \big[ \delta \hat{f}_{n^{\prime}} \,, \exp (i \chi_{n} \hat{X}_{30}) \hat{X}_{n}(\zeta_{n^{\prime}}) \big] \big\} \,. \label{26}
\end{align}

Using Eqs.~(\ref{25})--(\ref{26}) one can easily calculate the currents~$I_{\text{Q} |_{n}}$ and~$I_{n n^{\prime}}$. The expressions for~$I_{nn^{\prime}}$ have different forms depending on whether the matrices~$\hat{X}_{n}$ and~$\hat{X}_{n^{\prime}}$ correspond to different or equal chiralities.\cite{Moor_Volkov_Efetov_2015_d} We assume that the $n$\nobreakdash-th terminal corresponds to the $x$\nobreakdash-chirality, ${\hat{X}_{n} = \hat{X}_{x}}$ [see~Eq.(\ref{6})]. Then, for equal chiralities (${\hat{X}_{n} = \hat{X}_{n^{\prime}}}$) we obtain
\begin{align}
I_{nn^{\prime}}^{xx} = r_{n} r_{n^{\prime}} \sum_{\omega} F_{nn^{\prime}} \big[ &A (1 + \zeta_{n} \zeta_{n^{\prime}}) \sin \varphi_{nn^{\prime}} \label{27} \\
- &B (1 - \zeta_{n} \zeta_{n^{\prime}}) \sin \Phi_{nn^{\prime}} \big] \,, \notag
\end{align}
where ${F_{n n^{\prime}} = (2 \pi T) a f_{n-} f_{n^{\prime} -} \big/ \mathcal{E}^{3}}$, and ${\Phi_{n n^{\prime}} = \chi_{n} + \chi_{n^{\prime}} - 2 \chi_{\text{S}}}$, ${\varphi_{n n^{\prime}} = \chi_{n} - \chi_{n^{\prime}}}$.

If the triplet components at the $n$\nobreakdash-th and $n^{\prime}$\nobreakdash-th terminals correspond to different chiralities, e.g., ${\hat{X}_{n} = \hat{X}_{x}}$ and ${\hat{X}_{n^{\prime}} = \hat{X}_{y}}$ [see~Eq.(\ref{6'})], we find
\begin{align}
I_{nn^{\prime}}^{xy} = r_{n} r_{n^{\prime}} \sum_{\omega} F_{nn^{\prime}} \big[ &A (\zeta_{n} + \zeta_{n^{\prime}}) \cos \varphi_{nn^{\prime}} \label{27'} \\
- &B (\zeta_{n} - \zeta_{n^{\prime}}) \cos \Phi_{nn^{\prime}} \big] \,. \notag
\end{align}
With the help of Eqs.~(\ref{25'}),~(\ref{27}), and~(\ref{27'}) one can readily find the Josephson current through the $n$\nobreakdash-th terminal for an arbitrary multi-terminal structure with ``magnetic'' superconductors attached to a singlet superconductor via short normal wires.

Now, we discuss general properties of the partial currents~$I_{nn^{\prime}}^{xx}$ and~$I_{nn^{\prime}}^{xy}$, Eqs.~(\ref{27}) and~(\ref{27'}). In the absence of the singlet superconductor~S (in which case ${B = 0}$) only the first terms in Eqs.~(\ref{27}) and~(\ref{27'}) are finite. These terms turn to zero in the case of antiparallel filter axes (${\zeta_{n} = -\zeta_{n^{\prime}}}$). If these axes are parallel, the first term in Eq.~(\ref{27}) determines a usual Josephson current while the first term in Eq.~(\ref{27'}) describes a spontaneous current which exists in the absence of the phase difference and has a direction depending on the spin filter orientation~${\zeta_{n} = \pm 1}$.

In the presence of the superconductor~S (${B \neq 0}$) the second terms in Eqs.~(\ref{27}) and~(\ref{27'}) are not zero in the case of antiparallel filter axis. The second term in Eq.~(\ref{27}) is a ``Josephson''-like current~\cite{Moor_Volkov_Efetov_2015_d}, whereas the second term in Eq.~(\ref{27'}) determines a spontaneous current. The phase dependence of the second term in Eq.~(\ref{27}) at ${\zeta_{n} = -\zeta_{n^{\prime}}}$, ${I_{nn^{\prime}} \propto B ( 1 - \zeta_{n} \zeta_{n^{\prime}} ) \sin \Phi_{nn^{\prime}}}$, coincides with that in Ref.~\onlinecite{Liang_et_al_2011}, where the Josephson coupling occurred due to Majorana modes. We apply Eqs.~(\ref{27}) and~(\ref{27'}) to study the behavior of three- and four-terminal Josephson contacts.

\subsection{Three-terminal all-triplet S$_{\text{m}}$/Fl/n/Fl/S$_{\text{m}}$ junction}
\label{sec:four-terminal_triplet-only}

\begin{figure}
  \centering
  \includegraphics[width=1.0\columnwidth]{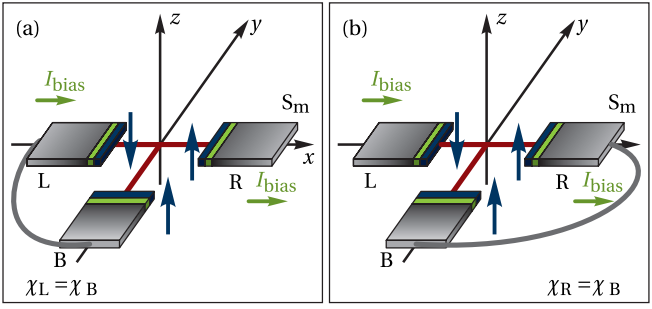}
  \caption{(Color online.) All-triplet three-terminal setup.}
  \label{fig:cross_triplet_only}
\end{figure}

First, we consider a simple case of a three-terminal Josephson junction consisting only of~S$_{\text{m}}$ superconductors, that is, only triplet components exist in the n~wire [cf.~Fig.~\ref{fig:System1a}~(a)]. In the case of parallel filter axes the behavior of the system under consideration is similar to that of an ordinary multi-terminal system consisting of singlet superconductors.\cite{Amin01,Amin02,Linder12,Mai13}

More interesting is the case when triplet Cooper pairs coming from the L\nobreakdash-reservoir and from B\nobreakdash- and R\nobreakdash-reservoirs (L, R, and B stand for, respectively, left, right, and bottom) have oppositely oriented spins. One can easily calculate the current through each interface from Eq.~(\ref{27}) taking into account that the coefficient~$B$ is zero, ${B = 0}$.

We consider two cases shown in Fig.~\ref{fig:cross_triplet_only}: (a)~the left S$_{\text{m}}$ is connected with the bottom S$_{\text{m}}$~superconductor so that ${\chi_{\text{L}} = \chi_{\text{B}}}$; and (b)~the right S$_{\text{m}}$ is connected with the bottom S$_{\text{m}}$~superconductor so that ${\chi_{\text{R}} = \chi_{\text{B}}}$.

\begin{enumerate}
\item[(a)] Using Eq.~(\ref{27}) and setting, for the sake of simplicity, ${F_{\text{RL}} = F_{\text{RB}} = F_{\text{LB}} \equiv F_{\text{0}}}$, i.e., all S$_{\text{m}}$~superconductors are identical (generalization to a more general case of different S$_{\text{m}}$ is trivial), we obtain for the setup displayed in Fig.~\ref{fig:cross_triplet_only}~(a):
\begin{align}
I_{\text{R}} &= \Big( r_{\text{R}} r_{\text{L}} \sum_{\omega \geq 0} F_{\text{0}}A \Big) \sin \varphi_{\text{RL}} = - I_{\text{B}} \label{27a} \,, \\
I_{\text{L}} &= I_{\text{R}} + I_{\text{B}} = 0 \,.  \label{27a'}
\end{align}
Equation~(\ref{27a'}) shows that no current flows through the left S$_{\text{m}}$~superconductor. The bias current ${I_{\text{bias}} = I_{R} =I_{\text{c}} \sin (\chi_{\text{R}} - \chi_{\text{B}})}$ flows through the usual  S$_{\text{mB}}$/n/S$_{\text{mR}}$ JJ, and the critical Josephson current~$I_{\text{c}}$ is given by the term in the square brackets. Note that all the currents~$I_{\text{L},\text{R},\text{B}}$ are the currents flowing from the n\nobreakdash-wire into a corresponding superconductor. This means that a negative~$I_{\text{B}}$ is the current which flows from the superconductor S$_{\text{mB}}$ into the n\nobreakdash-wire providing the continuity of the electric current.

\item[(b)] In this system, Fig.~\ref{fig:cross_triplet_only}~(b), the currents vanish,
\begin{equation}
I_{\text{R}} = I_{\text{B}} = I_{\text{L}} = 0 \,. \label{27b}
\end{equation}
In this case, the system is an insulator for dissipationless current because the right and bottom S$_{\text{m}}$~superconductors may be considered as a single spin-up S$_{\text{m}}$~superconductor which does not ``talk'' to the left spin-down superconductor~S$_{\text{mL}}$.
\end{enumerate}

\subsection{Four-terminal S$_{\text{m}}$/Fl/n/Fl/S$_{\text{m}}$ junctions
with a singlet superconductor}
\label{sec:four-terminal_with singlet}

\begin{figure}[tbp]
\centering 
\includegraphics[width=1.0\columnwidth]{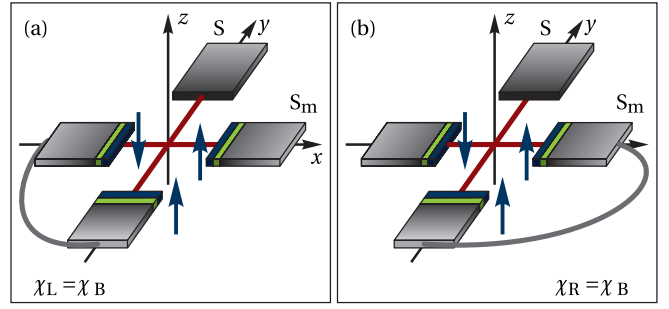}
\caption{(Color online.) Cross-geometry setup with a singlet superconductor
when (a)~bottom and left superconductors are connected (thus, ${\protect\chi%
_{\text{L}} = \protect\chi_{\text{B}}}$); (b)~bottom and right
superconductors are connected (thus, ${\protect\chi_{\text{R}} = \protect\chi%
_{\text{B}}}$). The blue arrows indicate the direction of filters which let
to pass triplet Cooper pairs with spins directed parallel to the particular
filter axis.}
\label{fig:cross_singlet}
\end{figure}

Next, consider the four-terminal JJ shown in Fig.~\ref{fig:cross_singlet}. It consists of a singlet superconductor~S and three~S$_{\text{m}}$ superconductors [right~(R), left~(L) and bottom~(B)] creating triplet components with equal chiralities (also, we assume ${\zeta _{\text{R}} = \zeta_{\text{B}} = -\zeta_{\text{L}} = 1}$) and connected by an n~wire. We consider again two cases: (a)~the phases of the left and bottom~S$_{\text{m}}$ superconductors are equal, ${\chi_{\text{L}} = \chi_{\text{B}}}$ (if in the loop shown in Fig.~\ref{fig:cross_singlet}, there is a magnetic flux~$\Phi_{H}$, then ${\chi_{\text{L}} = \chi_{\text{B}} + 2 \pi \Phi_{H} / \Phi_{0}}$, where~$\Phi_{0}$ is the magnetic flux quantum); and (b)~the phases of the right and bottom~S$_{\text{m}}$~superconductors are equal, ${\chi_{\text{R}} = \chi_{\text{B}}}$.

Using Eqs.~(\ref{26})--(\ref{27'}), we obtain for the Josephson currents through the right, bottom and left S$_{\text{m}}$/n~interfaces
\begin{align}
I_{\text{L}} &= -2 r_{\text{L}} \sum_{\omega \geq 0} F_{\text{0}} B \big[ r_{\text{R}} \sin \Phi_{\text{LR}} + r_{\text{B}} \sin \Phi_{\text{LB}} \big] \,,  \label{28''} \\
I_{\text{R}} &= r_{\text{R}} \sum_{\omega \geq 0} F_{\text{0}} \big[ A r_{\text{B}} \sin \varphi_{\text{RB}} - r_{\text{L}} B \sin \Phi_{\text{RL}} \big] \,,  \label{28} \\
I_{\text{B}} &= r_{\text{B}} \sum_{\omega \geq 0} F_{\text{0}} \big[ A r_{\text{R}} \sin \varphi_{\text{BR}} - r_{\text{L}} B \sin \Phi_{\text{BL}} \big] \,,  \label{28'}
\end{align}
where ${\varphi_{\text{BR}} = -\varphi_{\text{RB}} = \chi_{\text{B}} - \chi_{\text{R}}}$, ${\Phi_{\text{LR}} = \Phi_{\text{RL}} = \chi_{\text{R}} + \chi_{\text{L}} - 2 \chi_{\text{S}}}$, ${\Phi_{\text{LB}} = \Phi_{\text{BL}} = \chi_{\text{R}} + \chi_{\text{L}} - 2 \chi_{\text{S}}}$, and we assume again that ${F_{\text{XY}} = F_{0} = (2 \pi T) a f_{\text{X}-}f_{\text{Y}-} \mathcal{E}^{-3}}$.

One can see that ${I_{\text{L}} = I_{\text{R}} + I_{\text{B}}}$, i.e., in the applied approximation of a small S$_{\text{m}}$/n~interface transmittance, no current flows through the singlet S~superconductor although the phase of superconductor~S affects the currents~$I_{\text{R},\text{L},\text{B}}$.

\begin{figure}[tbp]
\centering 
\includegraphics[width=1.0\columnwidth]{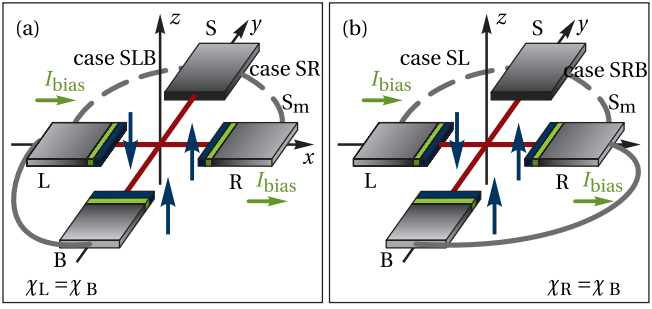}
\caption{(Color online.) Same cross-geometry setup with a singlet superconductor as in Fig.~\ref{fig:cross_singlet} but in addition, in each case the superconductor~S is connected to the left, respectively, right superconductor~S$_{\text{m}}$, thus four cases are considered (see main text): SLB, SR [in~(a)], SL, SRB [in~(b)].}
\label{fig:cross_singlet_additional}
\end{figure}

Consider furthermore the two different cases for an additional connection of the S~superconductor to the superconductors~S$_{\text{m}}$ (see Fig.~\ref{fig:cross_singlet_additional}), that is, for different relations between phases~$\chi_{\text{S}}$ and $\chi_{\text{R},\text{L},\text{B}}$.
\begin{enumerate}
\item[(a)~SLB] In the case~SLB, when there is an additional connection between the S~superconductor and the left S$_{\text{m}}$~superconductor [see Fig.~\ref{fig:cross_singlet_additional}~(a) with dashed line connecting S~and the \emph{left}~S$_{\text{m}}$ only], one has ${\chi_{\text{S}} = \chi_{\text{L}} = \chi_{\text{B}}}$. Then, we obtain ${I_{\text{bias}} = I_{\text{R}}}$ and
\begin{equation}
I_{\text{R}} = 2 r_{\text{R}} \sum_{\omega \geq 0} F_{0} \big[ A r_{\text{B}} - r_{\text{L}} B \big] \sin \varphi_{\text{RL}} \equiv I^{\text{(a)}}_{\text{c},\text{SLB}} \sin \varphi_{\text{RL}} \,.  \label{28a1}
\end{equation}
\item[(a)~SR] In the case~SR, when there is an additional connection between the S~superconductor and the right S$_{\text{m}}$~superconductor [see Fig.~\ref{fig:cross_singlet_additional}~(a) with dashed line connecting S~and the \emph{right}~S$_{\text{m}}$ only], one has ${\chi_{\text{S}} = \chi_{\text{R}}}$, while ${\chi_{\text{L}} = \chi_{\text{B}}}$. Then, the current~$I_{\text{R}}$ is
\begin{equation}
I_{\text{R}} = 2 r_{\text{R}} \sum_{\omega \geq 0} F_{0} \big[ A r_{\text{B}} + r_{\text{L}} B \big] \sin \varphi_{\text{RL}} \equiv I^{\text{(a)}}_{\text{c},\text{SR}} \sin \varphi_{\text{RL}} \,.  \label{28a2}
\end{equation}
Comparison of Eqs.~(\ref{28a1}) and~(\ref{28a2}) shows that although the singlet superconductor~S is electrically disconnected from the circuit (no current flows through the S/n~interface due to different symmetry of singlet and triplet components), it strongly affects the critical current of the system.

\item[(b)~SL] In the case~SL, when there is an additional connection between the S~superconductor and the left S$_{\text{m}}$~superconductor [see Fig.~\ref{fig:cross_singlet_additional}~(b) with dashed line connecting S~and the \emph{left}~S$_{\text{m}}$ only], one has ${\chi_{\text{S}} = \chi_{\text{L}}}$, while ${\chi_{\text{R}} = \chi_{\text{B}}}$. Then, we obtain ${I_{\text{bias}} = I_{\text{L}}}$ and
\begin{equation}
I_{\text{L}} = -2 r_{\text{R}} \sum_{\omega \geq 0} F_{0} B \big[ r_{\text{B}} + r_{\text{R}} \big] \sin \varphi_{\text{RL}} \equiv -I^{\text{(b)}}_{\text{c}} \sin \varphi_{\text{RL}} \,.  \label{28b1}
\end{equation}
\item[(b)~SRB] In the case~SRB, when there is an additional connection between the S~superconductor and the right S$_{\text{m}}$~superconductor [see Fig.~\ref{fig:cross_singlet_additional}~(b) with dashed line connecting S~and the \emph{right}~S$_{\text{m}}$ only], one has ${\chi_{\text{S}} = \chi_{\text{R}} = \chi_{\text{B}}}$. Then, the current~$I_{\text{R}}$ is
\begin{equation}
I_{\text{L}} = 2 r_{\text{R}} \sum_{\omega \geq 0}F_{0} B \big[ r_{\text{B}} + r_{\text{R}} \big] \sin \varphi_{\text{RL}} \equiv I^{\text{(b)}}_{\text{c}} \sin \varphi_{\text{RL}} \,. \label{28b2}
\end{equation}
We see that the critical current has different sign in these two cases.
\end{enumerate}

Note that the system in Fig.~\ref{fig:cross_singlet}~(b) corresponds to the ``magnetic'' case considered in Ref.~\onlinecite{Moor_Volkov_Efetov_2015_c} and the Josephson current is similar to that obtained in Ref.~\onlinecite{Moor_Volkov_Efetov_2015_d} for a three terminal Josephson junction. The case~(a) corresponds to ``nematic'' case since the spins of the triplet Cooper pairs coming from the~S$_{\text{L}}$ and~S$_{\text{B}}$ superconductors have opposite directions.

The case of different chiralities can be studied analogously.

\section{Conclusion.}

We have considered the dc Josephson effect in a diffusive multi-terminal Josephson junction which consists of some ``magnetic'' superconductors and one singlet superconductor connected via a normal wire~n. The ``magnetic'' superconductors are separated from the n~wire by spin filters so that only fully polarized triplet Cooper pairs can penetrate into the n~wire from these superconductors. We have shown that if the spin filters in the two-terminal S$_{\text{m}}$/n/S$_{\text{m}}$~Josephson junctions are antiparallel, there is no current at arbitrary S$_{\text{m}}$/n interface transparency. The presence of an additional s-wave singlet superconductor terminal~S results in a finite Josephson current flowing from the S~superconductor to the S$_{\text{m}}$~superconductors; one can speak of conversion of two singlet Cooper pairs into two triplet pairs with antiparallel total spins. The obtained unusual current--phase relation is compared with those which take place in other types of Josephson contacts,
for example, in JJs with the coupling due to Majorana modes.\cite{Liang_et_al_2011} Also, we have calculated the density of states in the normal wire for different types of two-terminal JJs and compared the DOS for nematic and magnetic cases.

A general formula for the Josephson current is derived for the case of multi-terminal JJs under assumption that the~S$_{\text{m}}$/n interface transparencies are small. We applied this formula for analysis of four-terminal JJs of different types. Both, nematic and magnetic cases,\cite{Moor_Volkov_Efetov_2015_c} can be realized with the aid of the considered four-terminal JJs. In a two-terminal structure with parallel filter orientations and in a three-terminal structure with antiparallel filter orientations of the ``magnetic'' superconductors with attached additional singlet superconductor, we find a nonmonotonic temperature dependence of the critical current. Also, in these structures, the critical current shows a dependence  on the exchange field in the ``magnetic'' superconductors with a Riedel-like peak. We analyzed also S$_{\text{m}}$/n JJs when only fully polarized triplet component exist in the n~wire.

We showed also that, in the applied first approximation in the transmission coefficient for S$_{\text{m}}$/n interfaces, no current flows through the singlet superconductor~S due to orthogonality of the triplet and singlet components, i.e., one can say that the superconductor~S is electrically disconnected from the circuit. Nevertheless, the phase~$\chi_{\text{S}}$ in the superconductor~S affects the Josephson current in the system.

All effects discussed above can be observed on systems that have been already studied experimentally.\cite{Keizer06,Aarts10,Aarts12,Birge10,Birge12,Zabel10,Petrashov06,Halasz_et_al_2011,Petrashov11,Blamire10,Blamire12,Leksin_et_al_2012,Gingrich_et_al_2012,Banerjee_et_al_2014} As spin filters, one can use either magnetic insulating layers or conducting magnetic half-metals.\cite{Keizer06,Aarts10,Aarts12} Some properties of ferromagnet/superconductor structures with a half-metallic layers have been analyzed in a recent publication.\cite{Mironov_Buzdin_2015} In particular, the authors of Ref.~\onlinecite{Mironov_Buzdin_2015} calculated the critical temperature~$T_{\text{c}}$ of S/F/HM~structures (where HM denotes a half-metal), which have been studied experimentally on MoGe/Cu/Ni/CrO$_2$~hybrids,\cite{Singh_et_al_2015} and analyzed the Josephson effect in S/F/HM/F/S~junctions. They have shown that a spontaneous phase difference ($\varphi$\nobreakdash-junction) arises in these junctions (a similar effect has been predicted in Refs.~\onlinecite{Eschrig03} and~\onlinecite{Moor_Volkov_Efetov_2015_c}).

Multiterminal JJs considered in the current Paper open new routs to vary the types of the current--phase relations for the Josephson current and to control the spin current.

\begin{acknowledgments}
We appreciate the financial support from the Deutsche Forschungsgemeinschaft via Projekt~EF~11/8\nobreakdash-2; K.~B.~E.~gratefully acknowledges the financial support of the Ministry of Education and Science of the Russian Federation in the framework of Increase Competitiveness Program of NUST~``MISiS'' (No.~K2\nobreakdash-2014\nobreakdash-015).
\end{acknowledgments}

\appendix

\section{Technical details on the derivation of density of states and of the current}
\label{app:technical_details}
\setcounter{paragraph}{0}

\paragraph{Antiparallel filter orientation (${\zeta_{\text{R}} = - \zeta_{\text{L}} \equiv \zeta}$).}

In this case, the matrix~$\hat{\Lambda}$ has the form ${\hat{\Lambda} = \hat{\Lambda}_{\text{n}} + \hat{\Lambda}_{\text{m}}}$ with
\begin{align}
\hat{\Lambda}_{\text{n}} &= 2 r_{\omega} \hat{X}_{30} \,, \label{8a} \\
\hat{\Lambda}_{\text{m}} &= 2 r_{\text{m}} \big[ g_{+} \hat{X}_{30} + f_{-} [ \cos(\varphi/2) \hat{X}_{11} - \zeta \sin (\varphi/2) \hat{X}_{12}] \big] \,.  \label{8}
\end{align}

The solution of Eq.~(\ref{4}) has the form
\begin{equation}
\hat{g} = \lambda_{\text{a}} \hat{\Lambda} \label{8'}
\end{equation}
with the constant~$\lambda_{\text{a}}$ which is found from the normalization condition, Eq.~(\ref{2}), as ${\lambda_{\text{a}}^{-1} = g_{\text{a}} \sqrt{1 + \gamma_{\text{a}}^{2}}}$, where
\begin{equation}
\gamma_{\text{a}} = r_{\text{m}} f_{-} g_{\text{a}}^{-1} \,, \label{eq:gamma_a_appendix}
\end{equation}
${g_{\text{a}} = r_{\omega} + r_{\text{m}} g_{+}}$, ${r_{\omega} = \omega / E_{\text{Th}}}$, and ${E_{\text{Th}} = D/L^{2}}$ is the Thouless energy; the functions~$f_{-}$ and~$g_{+}$ are defined in Eqs.~(\ref{eq:f_pm}) and~(\ref{eq:g_pm}).

The density of states is then calculated from Eqs.~(\ref{DOS1}) and~(\ref{8'}),
\begin{equation}
\nu(\varepsilon) = \Re\big[ (1 + \gamma_{\text{a}}^{2})^{-1/2}_{|_{\omega=-i \varepsilon}} \big] \,. \label{9_appendix}
\end{equation}
The DOS does not depend on the phase difference~${\varphi = \chi_{\text{R}} - \chi_{\text{L}}}$.

In order to find the Josephson current~$I_{\text{Q}}$, we use~Eq.~(\ref{8'}) and~Eq.~(\ref{I_Q}) with ${\hat{f}_{\text{R}} = [ \cos (\varphi/2) + i \sin (\varphi/2) \hat{X}_{30}] [\hat{X}_{11} - \zeta \hat{X}_{22}]}$. Simple calculations yield zero Josephson current
\begin{equation}
I_{\text{Q}} = 0 \,, \label{I_Qa_appendix}
\end{equation}
that is, in the considered system with antiparallel spin filter axes the Josephson current is zero at any transparencies of the S$_{\text{m}}$/n interfaces.

\paragraph{Parallel filter orientation (${\zeta_{\text{R}} = \zeta_{\text{L}} \equiv \zeta}$).}

In this case, we have for the matrix~$\hat{\Lambda}_{\text{m}}$
\begin{equation}
\hat{\Lambda}_{\text{m}} = r_{\text{m}} [g_{+} (\hat{X}_{30} + \zeta \hat{X}_{03} ) + f_{-} \cos (\varphi/2) \hat{X}_{\text{R(L)}} ] \,,  \label{10}
\end{equation}
where~$\hat{X}_{\text{R(L)}}$ is again one of the matrices defined in Eqs.~(\ref{6}) and~(\ref{6'}), and~${\varphi = \chi_{\text{R}} - \chi_{\text{L}}}$ is the phase difference. We look for a solution in the form
\begin{equation}
\hat{g} = a_{30} \hat{X}_{30} + \zeta a_{03} \hat{X}_{03} + a_{11} \hat{X}_{\text{R(L)}} \,,  \label{11}
\end{equation}
and write the matrix~$\hat{\Lambda}$ as
\begin{equation}
\hat{\Lambda} = G_{30} \hat{X}_{30} + \zeta G_{03} \hat{X}_{03} + F_{\text{m}} \hat{X}_{\text{m}} \,.  \label{12}
\end{equation}
where ${G_{30} = \tilde{g}_{\text{a}}}$, ${G_{03} = r_{\text{m}} g_{+}}$, and ${F_{\text{m}} = r_{\text{m}} f_{-} \cos(\varphi/2)}$.

From Eq.~(\ref{4}) we find ${a_{11} = \gamma_{\text{b}} (a_{30} + a_{03})}$ with
\begin{equation}
\gamma_{\text{b}}(\varphi) = F_{\text{m}} \cos (\varphi/2) g_{\text{b}}^{-1} \,, \label{13}
\end{equation}
where ${g_{\text{b}} = ( r_{\omega} + 2 r_{\text{m}} g_{+} )}$.

Next, the normalization condition yields ${a_{30}^{2} + a_{03}^{2} + 2 b^{2} = 1}$ and ${-a_{30} a_{03} = \zeta b^{2}}$. Introducing ${a_{30}^{2} + a_{03}^{2} = \varrho^{2}}$, ${a_{30} = \varrho \cos \alpha}$, and ${a_{30} = \varrho \sin \alpha}$, we obtain
\begin{align}
\sin (2 \alpha ) &= -\frac{2 \gamma_{\text{b}}^{2}}{1 + 2 \gamma_{\text{b}}^{2}} \,, \\
\varrho^{2} &= \frac{1 + 2 \gamma_{\text{b}}^{2}}{1 + 4 \gamma_{\text{b}}^{2}} \,. \label{15}
\end{align}
Thus, for the density of states, Eq.~(\ref{DOS1}), we find
\begin{align}
\nu(\varepsilon) &= \varrho \cos \alpha \label{16_appendix} \\
&= \frac{1}{2} \Re \big[ 1 + [1 + 4 \gamma_{\text{b}}^{2}(\varphi) ]^{-1/2}_{|_{\omega =-i \varepsilon}} \big] \notag
\end{align}
with~$\gamma_{\text{b}}$ defined in Eq.~(\ref{13}). The corresponding Josephson current is
\begin{equation}
I_{\text{Q}} = I_{\text{c}}(\varphi) \sin \varphi \,, \label{I_Qb_appendix}
\end{equation}
with the critical current
\begin{equation}
I_{\text{c}}(\varphi) = a r_{\text{m}}^{2}(2 \pi T) \sum_{\omega \geq 0} \frac{f_{-}^{2}}{\big(\frac{\omega}{\Delta} + 2 r_{\text{m}} \frac{E_{\text{Th}}}{\Delta} g_{+} \big) \sqrt{1 + 4 \gamma_{\text{b}}^{2}(\varphi)}} \,.  \label{I_c_b_appendix}
\end{equation}

One can see that the phase dependence of~$I_{\text{c}}$ leads to appearance of higher harmonics in the Josephson current, i.e., the current~$I_{\text{Q}}$ in Eq.~(\ref{I_Qb_appendix}) can be written as
\begin{equation}
I_{\text{Q}} = \sum_{n = 1}^{\infty} I_n \sin[(2 n + 1) \varphi] \,. \label{I_Qb_harmonics_appendix}
\end{equation}

\paragraph{Usual S/n/S Josephson junction.}

In this case, the matrix~$\hat{\Lambda}$ reads
\begin{equation}
\hat{\Lambda} = r_{\omega} \hat{X}_{30} + r_{\text{S}} [G_{\text{S}} \hat{X}_{03} + F_{\text{S}} \cos (\varphi/2) \hat{X}_{10}] \,,  \label{17}
\end{equation}
where ${F_{\text{S}} = \Delta/\sqrt{\omega^{2} + \Delta^{2}}}$. We choose ${\chi_{\text{R}} = -\chi_{\text{L}} = \varphi/2}$. The solution has a form similar to that for the case of antiparallel filter orientation, i.e., ${\hat{g} = \lambda_{\text{c}} \hat{\Lambda}}$, where ${\lambda_{\text{c}}^{-1} = \tilde{G}_{\text{S}} \sqrt{ 1 + \gamma_{\text{c}}^{2}}}$ and
\begin{equation}
\gamma_{\text{c}} = F_{\text{S}} \cos(\varphi/2) \tilde{G}_{\text{S}}^{-1} \label{eq:gamma_c_appendix}
\end{equation}
with $\tilde{G}_{\text{S}} = r_{\omega} + r_{\text{S}} G_{\text{S}}$.

The density of states is calculated as
\begin{equation}
\nu(\varepsilon) = \Re\big[ [1 + \gamma_{\text{c}}^{2}(\varphi)]^{-1/2}_{|_{\omega=-i \varepsilon}} \big] \,, \label{19_appendix}
\end{equation}
and the Josephson current reads
\begin{align}
I_{\text{J}} &= I_{\text{c}} \sin \varphi \,, \label{20_appendix} \\
\intertext{with}
I_{\text{c}} &= 2 a r_{\text{S}}^{2} (2 \pi T) \sum_{\omega \geq 0} \frac{F_{\text{S}}^{2}}{\sqrt{ 1 + \gamma_{\text{c}}^{2}(\varphi) }} \,. \label{20'_appendix}
\end{align}
The coefficient~$a$ is the same as in Eq.~(\ref{I_c_b_appendix}) and related with the critical current~$I_{\text{c}}$ via Eq.~(\ref{20_appendix}).


%

\end{document}